\documentstyle[12pt]{article}

\begin{document}

\begin{center}

{\Large \bf New possibilities for the gauging of chiral bosons}
\vspace{6mm}

{\Large E. M. C. Abreu$^{a,b}$\footnote{Present address: DCP, CBPF, Rua Xavier Sigaud 150, 22290-180, Urca, 
RJ, Brazil.}, 
A. de Souza Dutra$^{b}$ and 
C. Wotzasek$^{b,c}$\footnote{\sf E-mail: everton@cbpf.br; dutra@feg.unesp.br; clovis@if.ufrj.br and wotzasek@feg.unesp.br}}
\vspace{3mm}


\begin{large}
$^a$
{\it UNIFEI, Av.BPS 1303, Pinheirinho, 37500-903, Itajub\'a, MG, Brazil,} \\[0pt]
$^b$
{\it FEG/UNESP, PO Box 205, 12516-410, Guaratinguet\'a, SP, Brazil,}\\
$^c$
{\it IF, UFRJ, PO Box 68528, 21945-970, Rio de Janeiro, RJ, Brazil,}\\
\end{large}
\bigskip

\today


\end{center}



\begin{abstract} 
\noindent We study a new mechanism for the electromagnetic gauging of chiral bosons showing that new possibilities 
emerge for the interacting theory of chiral scalars.  We introduce a chirally coupled gauge 
field necessary to mod out the degree of freedom that obstructs gauge invariance in a system of two opposite chiral 
bosons soldering them together. PACS: 11.10.Ef, 04.65.+e
\end{abstract}
\bigskip

\newpage

{\bf 1.Introduction.}
In the last few years there has been a great amount of investigation on the proper way to 
gauge $2D$ self-dual fields [1-7]
and its $d$-dimensional extension \cite{eb,js}.
Despite of the successes of these indirect gauging schemes for the
Floreanini-Jackiw (FJ) model, the results
reported in \cite{gs} show clearly that the explicitly covariant model for chiral bosons put forward by Siegel \cite{siegel} also suffer from the same difficulties regarding
the coupling to gauge fields.  In this work we show how the results already reported in the literature can be obtained directly from first principles using the compatibility between field equations and chiral constraints as guiding rule.  We also show 
that, once this rule is relaxed, new and interesting possibilities emerge. These new gauging rules are the main  result of this paper.

In Section 2 we show the incompatibility between the gauge invariant constraint and
the field equations.   In Section 3 we introduce  
a gauge field, chirally 
coupled to the chiral matter, which is necessary in order to sold \cite{solda}
together a right and a left chiral boson.  
Our main result is presented in Section 4 where we use the dual projection scheme, 
proposed in \cite{dual,eu}, 
to study new gauging schemes.  Final discussions and perspectives are described in section 5.


{\bf 2.The gauging procedure.}
In the literature about this subject, the basic difficulty in the chiral boson gauging is pointed to be the loss 
of Lorentz covariance which refrain us from using the minimal substitution scheme. It is in fact easy to see that 
there is an incompatibility between the chiral constraint and the matter field equation.  
For instance, the action for a flat space-time free scalar field in the light-front 
coordinates (our notation: $x^{\pm}=
{1\over{\sqrt 2}}(x^0\pm x^1)$ ; $\epsilon^{+-}=\epsilon^{-+}=1$) is
${\cal L}= \partial_+\phi\,\partial_-\phi$.  The chiral constraint $\partial_-\phi\approx 0$ is consistent with 
the equations of motion (EM) before gauging, but not after.  Indeed,
suppose we gauge the system through the chiral derivative substitution
rule: $\partial_+\phi  \longrightarrow  \partial_+\phi$ and 
$\partial_-\phi  \longrightarrow  D_-\phi$
where $D_\pm\phi=\partial_\pm\phi + A_\pm$.  The EM
for the scalar field now reads $\partial_+D_-\phi \sim E$ with $E$ being the electric field on the line.  Notice that
the imposition of the gauge invariant constraint ($D_-\phi\approx 0$)
becomes inconsistent with the field equation due to the appearance
of the gauge anomaly.  Suppose that the equations of motion for some general chiral theory reads, before gauging, 
$L\:\partial_-\phi=0$,
with $L$ being the differential operator convenient for the Euler-Lagrange matter equation.  
If the gauging procedure chosen is, for instance, the direct substitution
of partial derivatives by their covariant counterparts, the outcome
after gauging will read $L\:D_-\phi\sim E$ displaying the above mentioned inconsistency.  For chiral bosons, 
the question one may ask is if there exists any gauging procedure in which the chiral
constraint remains compatible both before and after turning on the interactions.  As a matter of
fact, by examining the Lagrangian density proposed in \cite{bgp,harada}; 
${\cal L} = \partial_+\phi\,\partial_-\phi\,-\,\partial_-\phi\,\partial_-\phi +
2e\,(\partial_+\phi -\partial_-\phi) A_- \,+$ Gauge Terms; we get as the EM for the matter field,
$\partial_1\,D^e_-\phi=0$,
(where $D^{e}_{\pm}\,\phi=\partial_{\pm}+e\,A_{\pm}$) and this equation shows that the imposition of gauge 
invariant chiral constraint is not obstructed by the gauge anomaly.  
To obtain this result from first principles we need to find out what is the direct gauging scheme leading to this action. 
This we do next using the iterative Noether approach.
To keep the most generality in the formulation of the problem, we consider the case of a scalar field minimally coupled to a background gravitational field.   

The action for the standard minimal coupling of a scalar field to a metric $g_{\mu\nu}$ is
\begin{equation}
\label{geometric}
{\cal L}_0={1\over 2} \sqrt{-g} g^{\mu\nu}\partial_\mu\phi
\partial_\nu\phi\;\;.
\end{equation}
\noindent A convenient parametrization of the metric is given by
(observe that it does not correspond to a partial gauge-fixing but it 
is consequence of the Weyl symmetry)
\begin{equation}
\label{metric}
{1 \over 2}\sqrt{-g} g^{\mu\nu}=\lambda\,\left(
\begin{array}{cc}
{2\lambda_{--}} & {1+ \lambda_{++}\lambda_{--}}\\
{1+\lambda_{++}\lambda_{--}} & {2\lambda_{++}}
\end{array}
\right).\nonumber
\end{equation}
\noindent  where $\lambda=(1-\lambda_{++}\lambda_{--})^{-1}$.
For definiteness, from now on, let us consider the case of left chiral models.  
The (left) FJ and Siegel models can be obtained simply by truncation of this metric as
$\lambda_{--}=0$ for Siegel, and $\lambda_{--}=0, \lambda_{++}=-1$ for FJ, and this will be called as the 
{\it chiral limit}.
By computing the EM for the $\phi$ field we observe that
(before gauging) the chiral constraint cannot be imposed 
compatibly for arbitrary metric's components, differently from what happens in the flat space-time
case.  However, if we restrict ourselves to the (left) chiral model limit above, where
$\lambda_{--}=0$, 
then we have compatibility for the free theory ($L\:\partial_-\phi=0$ above) with the chiral constraint $\partial_- \phi = 0$. 
This is easily seen if we examine more closely the EM of (\ref{geometric}) before gauging, 
$0=\partial_+\{\lambda\,[2\lambda_{--}\partial_+\phi + (1+\lambda_{++}\lambda_{--})\partial_-\phi]\} 
+\partial_-\{\lambda\,[2\lambda_{++}\partial_-\phi + (1+\lambda_{++}\lambda_{--})\partial_+\phi]\}$.
The chiral constraint is not compatible with the EM.  
However, restriction to the {\it chiral limit} gives
$0 = (\partial_+ + \partial_-\lambda_{++} 
+\lambda_{++}\partial_-)\partial_-\phi$ for Siegel's EM and 
$0 = (\partial_+ - \partial_-)\partial_-\phi$ for FJ's EM, 
which shows compatibility with $\partial_-\phi\approx 0$.  After the Noether gauging
with (left) chiral currents (see below), the result is
$0 = (\partial_+ + \partial_-\lambda_{++} 
+\lambda_{++}\partial_-)D_-\phi$ for Siegel's EM and
$0 = (\partial_+ - \partial_-)D_-\phi$ for FJ's EM.
We see that for (left)
chiral couplings, the gauge invariant constraint can be imposed over the field equations
without being obstructed by the gauge anomaly.   This is
the only consistent possibility for the (left) chiral boson.

To implement the Noether procedure we need to compute the conserved currents associated to the global symmetries 
of the model. The axial current for (3) is
$J_{(A)}^+ = \lambda\,[2\lambda_{--}\partial_+\,\phi +(1+\lambda_{++}\lambda_{--})
\partial_-\,\phi]$ and 
$J_{(A)}^- =\lambda\,[2\lambda_{++}
\partial_-\,\phi +(1+\lambda_{++}\lambda_{--})\partial_+\,\phi]$.
Defining the vector current as dual to the axial current is not
really a restriction of our method since this is a feature of the
two dimensions.  In any case, one can show that in the non-Abelian case,
where the vector current can be defined as a Noether current, everything
works as discussed here.  Hence, the vector current is defined as dual to the axial one
$J_{(V)}^\mu=\mbox{}^*J^\mu_{(A)}$
where the usual Hodge transformation must be generalized to
$\mbox{}^*J^\mu_{(A)}=\sqrt{-g} g^{\mu\nu}\epsilon_{\nu\lambda}J_{(A)}^\lambda$ in order to take into account the 
presence of the gravitational background.  A simple calculation shows that
$J^+_{(V)}= -\partial_- \phi$ and
$J^-_{(V)}=\partial_+ \phi$
are topologically conserved, as it should.  Observe that
the vector current is metric independent, being the same for the chiral
models defined by truncation of the metric.

Having the axial and vectorial currents in hand, we are now in position to compute the right and left 
chiral currents.  We find,
$J^+_{(L)} = 2\lambda_{--}\,\lambda\,
\left(\lambda_{++}\partial_-\phi+\partial_+\phi\right)$ and
$J^-_{(L)} = 2\,\lambda\,
\left(\lambda_{++}\partial_-\phi+\partial_+\phi\right)$
for the left current, and
$J^-_{(R)} = 2\lambda_{++}\lambda\,
\left(\lambda_{--}\partial_+\phi+\partial_-\phi\right)$ and 
$J^+_{(R)} = 2\,\lambda\,
\left(\lambda_{--}\partial_+\phi+\partial_-\phi\right)$ for the right current.  Observe that the (left) {\it chiral} boson
{\it limit} ($\lambda_{--}=0$) kills the $J^+_{(L)}$ component, leaving
the left current holomorphically conserved, while the right chiral
current has both components non-vanishing.  This is certainly a
desired result for the chiral case.  In fact, in the flat space-time
theory for the free scalar field, there are two separated affine
invariances for the left and right chiral sectors.  These symmetries
are reflected by the fact that both the right and left chiral currents have
only one non-zero component, $J_{(L)}^-=J_{(L)}^-(x^+)$ and
$J_{(R)}^+=J_{(R)}^+(x^-)$, since 
$\partial_-J_{(L)}^- \,=\, 0$ and $\partial_+J_{(R)}^+ \,=\, 0$ 
and generate two commuting affine algebras.  However, in the
chiral case, only one of these currents keeps this property, i.e., either $J_{(L)}^- =J_{(L)}^-(x^+)$
but $J_{(R)}^+\neq J_{(R)}^+(x^-)$ or
$J_{(L)}^- \neq J_{(L)}^-(x^+)$ but $J_{(R)}^+ =J_{(R)}^+(x^-)$.  This can also be seen from the fact that, for chiral theories,
while the vector and axial transformations are global symmetries, the
affine transformations are semi-local symmetries.  Take for instance the
case of a left Siegel boson.  The semi-local shift $\phi\rightarrow
\phi +\alpha(x^+)$ certainly leaves the action invariant, but
$\phi\rightarrow \phi +\alpha(x^-)$ does not.  The Noether current is
immediately identified as $J^- =2(\partial_+\phi+\lambda_{++}\partial_-\phi)$ and $J^+=0$
which is easily seen to be the (left) {\it chiral limit} of $J^+_{(L)}$ and $J^-_{(L)}$ above.

We shall examine next the coupling of all these four currents with an external electromagnetic field.    We do this iteratively,
introducing the necessary Noether counter-terms.  Let us first examine the
coupling with the vector current $J^+_{(V)}$ and $J^-_{(V)}$ in the standard way:
${\cal L}_0\rightarrow {\cal L}_1 = {\cal L}_0 + A_+J^+_{(V)}+A_-J^-_{(V)}$,
where ${\cal L}_0$ is defined by Eq. (\ref{geometric}).  The global vector symmetry present in ${\cal L}_0$ has been lift to a local symmetry in ${\cal L}_1$. We observe
that, after gauging, the vector current remains conserved, but
the axial current does not. A direct calculation shows that
$\partial_\mu J^\mu_{(A)}=\partial_-A_+-\partial_+A_-=E$
which, again, being independent of the metric elements,
is valid for all cases.  By computing the EM for ${\cal L}_1$ we clearly see that the original compatibility 
between constraint and EM has been destroyed for the gauging with vector
currents  $J^+_{(V)}$ and $J^-_{(V)}$ due to the presence of the anomaly.  Therefore, we have
to rule out the vector current gauging as inappropriate for chiral theories.

Let us consider next the case of axial coupling.  The free action changes to 
${\cal L}_0\rightarrow{\cal L}_1={\cal L}_0 + A_+J^+_{(A)}+A_-J^-_{(A)}$.
\noindent Differently from the vector case, under an axial transformation the action ${\cal L}_1$ does not remain 
invariant but its variation is given by
$\delta{\cal L}_1\,=\,-\,\delta \{\lambda\,[ \lambda_{--} A_+^2 + \lambda_{++} A_-^2 
+{1+\lambda_{++}\lambda_{--}}A_+A_- ] \}$.  A further modification of the action as
\begin{equation}
{\cal L}_1\rightarrow{\cal L}_2\,=\,{\cal L}_1 + {1
\over{1-\lambda_{++}\lambda_{--}}}\left[\lambda_{--}A_+^2+\lambda_{++}A^2_-  
\,+\,  \left(1+\lambda_{++}\lambda_{--}\right)A_+A_-\right]
\end{equation}

\noindent leaves the action ${\cal L}_2$ invariant under an axial transformation. 
It is simple to check that here, the axial current
remains conserved, while the gauge coupling modifies the vector current,
which now fails to be conserved.  Finally, by checking the
EM, one notices the incompatibility of the gauged constraint with the
field equation, which also rules out this coupling as inappropriate for the chiral models.

We are then left only with the possibilities of chiral current couplings. 
We have explicitly checked that for left chiral bosons, the coupling with the
right chiral current results incompatible.  Now, let us work out explicitly
the coupling of the left chiral current and verify
that this is the only possible consistent way of coupling.  The Noether
procedure then gives 
${\cal L}_0\rightarrow{\cal L}_1={\cal L}_0 + A_+J^+_{(L)}+A_-J^-_{(L)}$ whose variation reads 
$\delta{\cal L}_1 \,=\, {\lambda}[(2\lambda_{++}\lambda_{--}A_+\partial_-\alpha + 2A_-
\partial_+\alpha) -\delta (\lambda_{--}A_+^2+\lambda_{++}A_-^2)]$.
The second term can be reabsorbed into a redefinition of
the action as
\begin{equation}
\label{axvar}
{\cal L}_1\rightarrow{\cal L}_2={\cal L}_1 +\lambda_{++}A_-^2 +
\lambda_{--}A_+^2\;\;, 
\end{equation}

\noindent but the first piece cannot be eliminated by any choice of
a Noether counter-term.  This is true even for the truncated chiral
limit but it is hardly a surprise since the gauged action for chiral models are
not expected to be gauge invariant.  However, this action
has the nice property of having its variance independent of the 
matter fields.

The important point to observe is that the truncation process
used to go from the non-chiral to the chiral case does not change the
nature of the coupling, which means that the vector, axial and chiral
couplings studied above remains the same after truncation.  This is
certainly different from the projection process using chiral constraints
that transform, for instance, the vector coupling into a chiral
coupling. 

It must be pointed out that the FJ chiral limit of ${\cal L}_2$ in Eq. (\ref{axvar}) is identical to that in  
Refs. \cite{bgp,harada}, obtained via left projection of the chiral Schwinger model (SM), but the other limits here are 
new results.  In particular we emphasize that we did not invoke the chiral SM at any level, but derive 
our result from first principles.

Finally, we believe that for chiral bosons actions which use infinite auxiliary fields \cite{mwyclovis} (the equivalence
between both proposals was demonstrated in Ref. \cite{clovis}), 
the procedure is the same, but now we have a sum of infinite chiral currents, one for each field, 
that are embedded 
in an external electromagnetic field.  Similarly, for the PST formulation \cite{pst} of the Siegel chiral boson, 
it is not difficult to see that that procedure also works.

{\bf 3. A gauging example.}
As a simple illustration of the use of the gauged theory,
the property mentioned above is explored now in order to sold
together two bosons of opposite chiralities. Details can be found in Ref. \cite{solda}.  
The soldering process is non trivial since
the simple sum of the actions of a right and a left chiral boson is not equal
to the action of a single scalar field.   

For computational convenience we use front-form variables. 
In this coordinate system, the action for left and right FJ chiral
bosons read, respectively 
${\cal L}_0^{(\pm)} = \mp\dot\phi_{\pm}\phi_{\pm}^\prime -
(\phi_{\pm}^\prime)^2$, where dot and prime have their usual significance as time and
space derivatives respectively.  We know, from their field equations, that these
models have a residual
invariance under a semi-local transformation 
$\phi_{\pm}\rightarrow\tilde\phi_{\pm}=\phi_{\pm} +\alpha_{\pm}(t)$.  Therefore, if one defines a scalar field as a 
combination of these chiral ones as $\Phi = \phi_+ - \phi_-$, then clearly the combination of the two semi-local
transformations above will not lead to a vector transformation for the
scalar field, unless some constraint is imposed over each individual
component.  This is the role played by the gauge field.  We can then
follow the gauging procedure described above to obtain the
action of an interacting chiral boson coupled to a gauge field through
their chiral left and right currents, respectively
${\cal L}_0^{\pm}\rightarrow {\cal L}_1^{\pm}={\cal L}_0^{\pm}
\:\mp 2A\left(\dot\phi_{\pm}\pm \phi_{\pm}^\prime\right)$.
Here $A$ is the space-component of the gauge field, i.e., $A$ is the so-called soldering field, an auxiliary field that
will be eliminated (integrated) through the field equations (or through the path integral, as below).  
Although
each individual (gauged) action is variant under a gauge
transformation, as we have seen in the last section, (see Eq. \ref{axvar})
one can verify that the new action 
${\cal L}={\cal L}_1^- + {\cal L}_1^+ - 2 A^2$ is indeed invariant.  The last term is a contact term that compensates for the non-invariances of each chirality.  Now, we can integrate out the
gauge field $A$, as
$e^{i\int d^2 x {\cal W}}=\int\, [dA]\,e^{i\int d^2 x {\cal L}}$
to obtain ${\cal W}={1\over 2}\partial_\mu\Phi\partial^\mu\Phi$, 
which is the action for the scalar field $\Phi=\phi_+\,-\,\phi_-$ defined as a combination of a right and a 
left chiral boson.


{\bf 4. The dual projection approach.} In this section we consider another approach to the chiral gauge problem based on
the dual projection \cite{dual,eu} of scalar fields.    The basic idea in this scheme is to gauge the scalar 
theory, using the obvious covariance that it possess, and then ``break" the basic fields, in phase space, into 
its chiral components.  For the definiteness, let us consider a well known generalized chiral SM (GCSM),
${\cal L} = \bar \Psi \gamma^\mu\left(i\partial_\mu + e_R A_\mu \frac {1+\gamma_5}2 + 
e_L A_\mu \frac{1-\gamma_5}2\right)\Psi$
\noindent where $\Psi$ is a two-component spinor
$\Psi = \left(
\begin{array}{c}
\psi_R\\
\psi_L
\end{array}\right)$,
and $\psi_L$ and $\psi_R$ are the Weyl components.   
Its quantum dynamics is easily computed using the bosonized version after a convenient redefinition of the fields,
${\cal L}_{Boson}={1\over 2}\partial_\mu\phi\partial^\mu\phi -  e_- A_\mu \partial^\mu \phi
+ e_+ A_\mu \epsilon^{\mu\nu}\partial_\nu \phi + g^2 \frac{a}{2} A_\mu A^\mu$
where  $e_R \,=\, \frac 12 \left(e_+ + e_-\right)$, $e_L \,=\, \frac 12 \left(e_+ - e_-\right)$, 
$g^2 = (e_-^2 + e_+^2)/2$ and $a$ is an arbitrary regularization parameter.
The interesting point in working with the GCSM is that one can easily obtain the four versions of 
the SM considered in the literature: Vector SM ($e_- = 0 \Rightarrow e_L=e_R=e$);
Axial SM ($e_+ = 0 \Rightarrow e_L= - e_R=e$);
Right Chiral SM ($e_L = 0 \Rightarrow e_+=e_-=e$) and
Left Chiral SM ($e_R = 0 \Rightarrow e_+= - e_-=e$).
The dual projection scheme permits one to realize that the second-order differential EM for the 
matter field coming from action ${\cal L}_{Boson}$ contains both right and left moving solutions. 
Let us reduce this action to its first-order form as,
\begin{equation}
\label{firstorder}
{\cal L}\,=\,\pi \dot \phi - \frac 12 \pi^2 - \frac 12 \left(\phi '\right)^2 - 
\left(e_- A_0 + e_+ A_1\right) \pi  
\,+\, \left(e_- A_1 + e_+ A_0\right)  \phi'
\,+\, \frac 12 A_\mu M^{\mu\nu}A_\nu
\end{equation}
where $A_\mu M^{\mu\nu}A_\nu= ag^2 \; A_\mu A^\mu - \left(e_- A_0 + e_+ A_1\right)^2$  
and $\pi$ is the second field mentioned above obtained through Legendre transformation from ${\cal L}_{Boson}$. 
We observe that the fields $\phi$ and $\pi$ cannot be our chiral fields describing, independently, the right and 
left dynamics since they appeared coupled in the symplectic sector of Eq. (\ref{firstorder}).  We must therefore look 
for a linear combination of them that diagonalize the action.  If we introduce the new fields $\varphi$ and $\rho$ 
as an $SO(2)$ rotation of the $\pi$ and $\phi$ fields as,
$\left(
\begin{array}{c}
\pi \\
 \phi '
\end{array}
\right)=\left(
\begin{array}{cc}
\cos \theta & \sin \theta\\
- \sin\theta & \cos\theta
\end{array}
\right)
\left(
\begin{array}{c}
\varphi ' \\
 \rho '
\end{array}
\right)$ 
then we quickly realize that, in order to diagonalize the action (\ref{firstorder}), we must have $\theta = \pi/4 $, 
and consequently the following canonical transformations (CT) \cite{bg}
$\phi '\,=\, \varphi ' + \rho '$ and 
$\pi \,=\, \varphi ' - \rho '$,
which we take as our definition of the chiral fields $\varphi$ and $\rho$.
In terms of these new fields the action for the GCSM, ${\cal L}_{Boson}$, now reads, 
${\cal L}_{Boson} = {\cal L}_\varphi + {\cal L}_\rho + {\cal L}_A$
where ${\cal L}_\varphi= \dot\varphi \varphi ' -{\varphi '}^2 + 2 e_L \,\varphi '\, A_-$; 
${\cal L}_\rho= - \dot\rho \rho ' -{\rho '}^2 + 2 e_R \,\rho '\, A_+$ and
${\cal L}_A = \frac 12 A_\mu M^{\mu\nu} A_\nu$ with $A_\pm = A_0 \pm A_1$.

There are many interesting points to observe in this now diagonalized action. Notice that in the free case the chiral 
fields satisfy $\partial_-\phi=0$ and $\partial_+\rho=0$ so that they are, respectively, left and right chiral modes. 
The action for the free components are those proposed by FJ. Also, no matter which model is in 
discussion, the interaction piece shows that these fields only couple to a proper chiral combination of the electromagnetic 
field, $A_-$ and $A_+$ respectively, but the coupling constant is, obviously, model dependent.  For the vector and axial 
SMs both chiral components are coupled to the photon field but the chiral SM will leave one 
chirality free, either $\varphi$ when $e_L=0$ or $\rho$ when $e_R=0$. In all these cases the mass term will change 
appropriately.
What should be regarded as the most important point is that this dual projection shows the chiral bosons correctly 
coupled to the photon field as proposed by Refs. \cite{bgp,harada} .
However in our case we got the coupling of both chiralities at once.  In the name of completeness, 
analysing the case of chiral
bosons's models with infinite auxiliary terms coupled to gravity, using the dual projection, it was shown 
\cite{bw}
that this model has nonmovers fields, the notons (introduced by Hull \cite{hull} to cancel the Siegel anomaly), 
which
couple naturally with the gravitational field.  For the PST version of Siegel chiral boson, it was proved that its
spectrum is composed by a FJ particle and a PST version of the noton \cite{eu}.

We are now in position to compare this result with the chiral projection technique adopted in Ref. \cite{harada}.  
From the CT defined above we get $\pi -\phi'\,=\, - 2 \rho'$ and $\pi + \phi'\,=\, 2 \varphi$, 
so that the chiral constraints $\pi\, \pm\, \phi '=0$ lead to $\varphi = 0$ and $\rho=0$ as constraints, 
respectively.  
It should be observed that the chiral component eliminated through this process need not be a free component, so that one 
does not need to start with a chiral SM in order to get the gauged chiral boson. So, except for some very 
awkward possibilities, any model will do.
Finally, we must observe that the CT above leads to a very sensible interpretation of the chiral bosons 
$\varphi$ and $\rho$ as being the bosonized components of the Weyl fermions in the GCSM since they are coupled 
to the gauge field with the same strength as their fermionic counterparts.   


{\bf 5. Final Discussions.} 
In conclusion, in this work we have studied the problem of coupling self-dual scalar
fields in $2D$ to an external electromagnetic field described by a vector
potential $A_\mu$.  We have recognized the basic difficulty
as an incompatibility between the gauge invariant
chiral constraint and the field equation for the matter field.  Using
consistency as a guiding rule, we have worked out the
coupling of gauge fields with different matter currents and, observed
that the only consistent coupling for a left chiral matter is
with a left chiral current.  This explain the results obtained  with the
use of the chiral projector on the SM.  As an application
we verified that in order to sold together a left and a right chiral
boson into a scalar field, we must (chirally) couple them to a gauge field that
will mod out the degree of freedom that obstructs gauge
invariance.
 

{\bf 6. Acknowledgments.} 
This work was partially supported by Funda\c{c}\~ao de Amparo \`a Pesquisa
do Estado de S\~ao Paulo (FAPESP) and Conselho Nacional de Desenvolvimento Cient\'{\i}fico e Tecnol\'ogico (CNPq).
  CNPq and  FAPESP are Brazilian research agencies.

\end{document}